\documentstyle[emulateapj_accepted,apjfonts,epsf]{article}

\def\ltsima{$\; \buildrel < \over \sim \;$}
\def\gtsima{$\; \buildrel > \over \sim \;$}
\def\lsim{\lower.5ex\hbox{\ltsima}}
\def\gsim{\lower.5ex\hbox{\gtsima}}
\newcommand{\etal}{et al.~}
\newcommand{\PSRI}{J1701$-$30}
\newcommand{\PSRII}{J1740$-$53}
\newcommand{\PSRIII}{J1807$-$24}
\newcommand{\PSRIV}{J1910$-$59}
\newcommand{\GCI}{6266}
\newcommand{\GCII}{6397}
\newcommand{\GCIII}{6544}
\newcommand{\GCIV}{6752}
\newdimen\digitwidth    %define ! a one digit width for tables
\setbox0=\hbox{\rm0}
\digitwidth=\wd0
\catcode`!=\active
\def!{\kern\digitwidth}
\renewcommand{\today}{October 12, 2000}
\begin{document}

\title{
\vskip -1.0truecm
Discovery of short-period binary millisecond pulsars in four globular 
clusters}

\author{
N.~D'Amico,\altaffilmark{1,2}
A.~G.~Lyne,\altaffilmark{3}
R. N. Manchester,\altaffilmark{4} 
A.~Possenti,\altaffilmark{1}
and F.~Camilo\altaffilmark{5}}
\medskip

\affil{\altaffilmark{1}Osservatorio Astronomico di Bologna,
Via Ranzani 1, 40127 Bologna, Italy; damico@bo.astro.it}
\affil{\altaffilmark{2}Istituto di Radioastronomia del CNR,
Via Gobetti 101, 40126 Bologna, Italy}
\affil{\altaffilmark{3}University of Manchester, Jodrell Bank
Observatory, Macclesfield, SK11~9DL, UK}
\affil{\altaffilmark{4}Australia Telescope National Facility,
CSIRO, PO Box 76, Epping, NSW 2121, Australia}
\affil{\altaffilmark{5}Columbia Astrophysics
Laboratory, Columbia University, 550 West 120th Street, New York, 
NY 10027}

\bigskip

\begin{abstract}

We report the discovery using the Parkes radio telescope of binary
millisecond pulsars in four clusters for which no associated 
pulsars were previously known.  The four pulsars have pulse periods 
lying between 3 and 6 ms. 
All are in circular orbits with low-mass companions and have orbital 
periods of a few days or less. One is in 
a 1.7-hour orbit with a companion of planetary mass. Another is eclipsed by 
a wind from its companion for 40\% of the binary period despite being 
in a relatively wide orbit.  These discoveries result from the use of improved
technologies and prove that many millisecond pulsars remain to be found in 
globular clusters.

\end{abstract}


\section{Introduction}

Globular clusters are a rich source of millisecond pulsars. Besides the
evolution of primordial binary systems, exchange interactions in their core
result in the formation of binary systems containing neutron stars. These
systems subsequently evolve, spinning up the neutron star through mass
accretion (\cite{sb76}; \cite{bv91}; \cite{ka96}).  The millisecond pulsars
formed in this way are among the most stable clocks in nature and are
valuable for studies of the dynamics of clusters, the evolution of binaries
embedded in them, and the interstellar medium (\cite{phi92}; \cite{hmg+92};
\cite{bv91b}; \cite{fcl+00}).  However, they are difficult to find because
the pulsed emission is weak and distorted by propagation through the
interstellar medium, and the apparent pulse period may change rapidly
because of binary motion.  We have constructed a new high resolution
filterbank system to improve the sensitivity to pulsars with high dispersion
measure (DM) and we have implemented a new multi-dimensional code to search
over a range of DMs and over a range of accelerations resulting from binary
motion.  Using these new capabilities, we have undertaken a search of
globular clusters for millisecond pulsars. So far, we have discovered four
millisecond pulsars in four clusters, none of which had previously known
pulsars associated with them. All of these pulsars are members of
short-period binary systems, and two of them have relatively high DM values.
These detections bring the number of clusters containing known pulsars to
16.  

\section{Observations and results}

The new discoveries were made during a search of globular clusters using the
Parkes 64-m radio telescope and the dual-polarization center beam of the
multibeam receiver (\cite{swb+96}) centered at 1374 MHz. With a system
temperature of $\sim 21$~K and bandwidth of 256 MHz, this receiver has very
high sensitivity. The effects of interstellar dispersion were removed by
using a filterbank having $512\times0.5$-MHz contiguous frequency channels
for each polarization.  This new filterbank system has sufficient resolution
to allow detection of millisecond pulsars with DMs of more than 200
cm$^{-3}$ pc at frequencies around 1400 MHz. Signals from individual
channels are detected, added in polarization pairs, high-pass filtered,
integrated and 1-bit digitized every 125 $\mu$s, and then
recorded on magnetic tape for off-line analysis.  With an integration time
of up to 2.3 hours per cluster, the nominal ($8\sigma$) sensitivity  to a
typical 3 ms pulsar with DM $\sim$ 200 cm$^{-3}$ pc is about 0.14 mJy,
several times better than previous searches.  The half-power width of the
telescope beam at 1374 MHz is 14 arcmin, encompassing most or all of the
mass of the target clusters.

Off-line processing is performed on a cluster of Alpha-500 CPUs at the
Osservatorio Astronomico di Bologna or on the Cray T3E multiprocessor system
of the CINECA Supercomputing Center, near Bologna, Italy. Each data stream
is split into non-overlapping segments of 2100, 4200 or 8400 sec and these
are separately processed. Data are first de-dispersed over a wide range of
dispersion measures centered on the value expected for each cluster on the
basis of a model of the Galactic electron layer (\cite{tc93}). Two different
search algorithms were used at different phases of the search. Initially,
the time domain data were interpolated to compensate for an acceleration and
then transformed using a Fast Fourier Transform (FFT), with many trials to
cover the expected acceleration range (cf. Camilo et al. 2000). Since this
analysis involves many FFTs, it is relatively slow. Later analyses exploited
the fact that even highly accelerated binaries have significant spectral
power in a zero-acceleration FFT. Time-domain data were fast-folded at
periods corresponding to spectral features above a threshold to form a
series of `sub-integration arrays' and these arrays were searched for the
parabolic signatures of an accelerated periodicity.  Parameters for final
pulse profiles having a signal-to-noise ratio above 8 were output for visual
examination.

\begin{figure*}
\vskip 1.0truecm
\epsfxsize=18truecm
\epsfysize=5truecm
\epsfbox{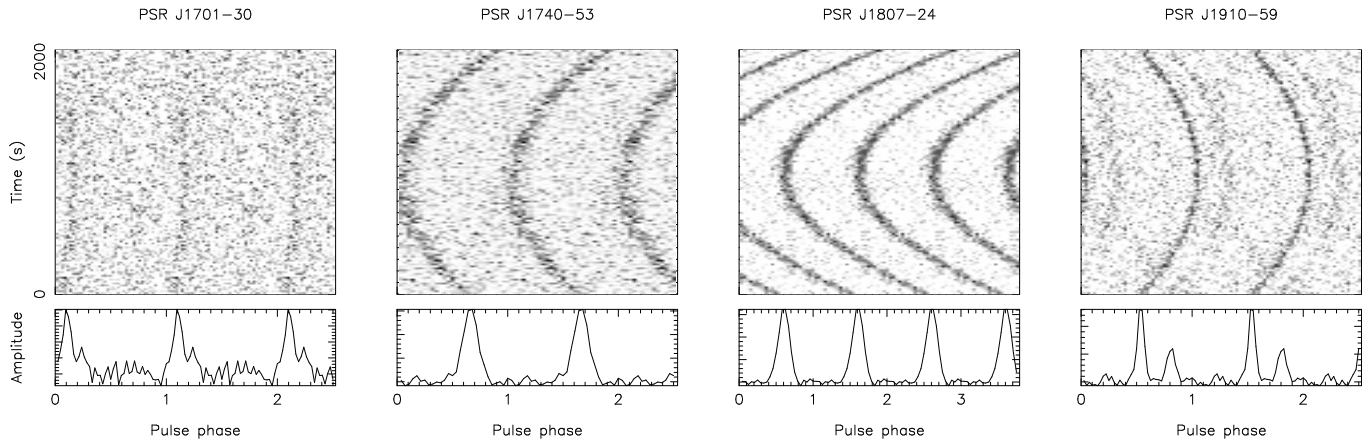}
\vskip -0.5truecm
\figcaption[fig1.ps]{\label{fig:fig1} 
\small{Time-resolved (upper panels) and
integrated (lower panels) pulse profiles for the four new binary millisecond
pulsars over a 2100-sec observation. Each horizontal line in the upper
panels represents the mean pulse profile for 16 seconds of data folded with
the mean period value indicated by the spectral analysis.  A pulsar with a
stationary period would form a straight trace in this diagram.  The
integrated pulse profile for PSR J1910$-$59 has two widely separated
components. PSR J1701$-$30 may have significant emission in the interpulse
region.}}
\end{figure*}

Figure 1 shows time-resolved and average pulse profiles for the
four new pulsars.  The time-resolved diagrams shown in the top panels of the
figure are the sub-integration arrays used in the fast folding.  Parameters
for the four pulsars and the associated globular clusters are given in Table
1.  For PSR J1807$-$24, these result from a coherent phase fit
using the timing program
TEMPO\footnote{http://pulsar.princeton.edu/tempo}. Parameters for PSRs
J1701$-$30 and J1740$-$53 are derived by fitting observed periods obtained
from short observations (typically 30 min) with a binary model. For PSR
J1910$-$59, the values are obtained by fitting to the period and
acceleration values measured at various epochs (see \S 2.4).
Uncertainties are given in parentheses and refer to the last quoted
digit. Except for PSR J1910$-$59, the quoted epoch is the time of ascending
node of the binary motion. The mass function, $f(M_p) = M_c^3 \sin^3 i /
(M_p + M_c)^2$, where $M_p$ and $M_c$ are the pulsar and companion masses,
is computed from the observed binary parameters. A minimum mass for the
companion may be derived from this by assuming an edge-on orbit, i.e.,
$i=90^\circ$, and a pulsar mass $M_p = 1.35$ M$_\odot$. Flux densities are
approximate only and are estimated from the system sensitivity and observed
signal-to-noise ratio. PSR J1740$-$53 is eclipsed for part of its orbit, and
the quoted flux density refers to epochs away from eclipse. PSR J1910$-$59
scintillates markedly and the quoted flux density is an estimated mean value
including the non-detections.  Cluster parameters are from the compilation
of Harris (1996)\nocite{har96}\footnote{see update at
http://physun.mcmaster.ca/Globular.html}. The telescope beam was centered on
the nominal cluster position. $z$ is the perpendicular distance from the
Galactic plane, and DM sin $|b|$, where $b$ is the Galactic latitude, is the
`$z$- component' of dispersion measure. 

\begin{deluxetable}{lllll}
\scriptsize
%\tablewidth 1.95\columnwidth
\tablecaption{\label{tab:params} Observed and derived parameters for four
millisecond pulsars and their associated clusters}
\tablecolumns{5}
\tablehead{\colhead{Pulsar}&\colhead{PSR {\PSRI}}
&\colhead{PSR {\PSRII}}&\colhead{PSR {\PSRIII}}&\colhead{PSR {\PSRIV}}}
\startdata
Period  (ms)      & 5.2415660(4) & 3.6503298(3) &3.0594487974(3) &
3.266182(3) \\
Epoch (MJD)            & 51698.415(2) & 51615.658(4) & 51734.97510(1)&
51745.0\\
DM (cm$^{-3}$ pc) & 114.4(3) & 71.8(2) & 134.0(4) & 34(1) \\
Orbital period (days)  & 3.805(1) & 1.3541(1) & 0.071092(1) & 0.865(5)\\
a ${\rm sin}i$ (light-s)      & 3.48(1) & 1.66(2) & 0.01220(3) & 1.27(4)\\
Mass function (M$_\odot$) & 0.0031& 0.0027 & 3.85$\times$10$^{-7}$ & 0.0029 \\
Minimum companion Mass (M$_\odot$) & 0.19 & 0.18 & 0.009 & 0.19\\
Flux density (mJy) & 0.2 & 1.0 & 1.3 & 0.2\\
Luminosity (mJy kpc$^2$) & 10  & 5 & 8 & 3 \\ 
DM sin $|b|$ (cm$^{-3}$ pc) & 14.7 & 14.8 & 4.9 & 14.4 \\
\tableline
$~~~~~~~~~~~~~~~~~$Cluster & NGC {\GCI}  & NGC {\GCII} & NGC {\GCIII}  &
NGC {\GCIV}\\
\tableline
R. A. (J2000) 
& $17^{\rm h} 01^{\rm m} 12\fs6$
& $17^{\rm h} 40^{\rm m} 41\fs3$
& $18^{\rm h} 07^{\rm m} 20\fs6$
& $19^{\rm h} 10^{\rm m} 51\fs8$ \\
Dec.  (J2000) 
& $-30^\circ 06^\prime 44^{\prime\prime}$
& $-53^\circ 40^\prime 25^{\prime\prime}$ 
& $-24^\circ 59^\prime 51^{\prime\prime}$
& $-59^\circ 58^\prime 55^{\prime\prime}$ \\
Galactic longitude & 353.6 & 338.2 & 5.8 & 336.5 \\
Galactic latitude & $+$7.3 & $-$11.9 & $-$2.2 & $-$25.6 \\
Distance (kpc) & 6.7  & 2.2 & 2.5 & 3.9 \\ 
$z$ (kpc) & $+$0.85  & $-$0.45 & $-$0.10 & $-$1.7 \\ 
Log(L$_{0}$)(L$_{\odot}$ pc$^{-3}$) & 5.15 & 5.69 & 5.78 & 4.92 \\
r$_{\rm core}$ (arcmin) & 0.18 & 0.05 & 0.05 & 0.17 \\
r$_{\rm tidal}$ (arcmin) & 9 & 16 & 2.1 & 55 \\
\enddata

\end{deluxetable}

The four pulsars have pulse periods lying between 3 and 6 ms and 
all are in circular orbits with relatively low-mass companions.  Two of
them (PSRs J1701$-$30 and J1807$-$24) have relatively high DM and would
not be detectable without the use of the new filterbank system.  The orbital 
periods range from 3.8 days down to only 1.7 hours for PSR J1807$-$24.  Three
of them (PSRs J1740$-$53, J1807$-$24 and J1910$-$59) show significant
acceleration over a relatively short time scale. While PSRs J1740$-$53 
and J1807$-$24
are relatively strong,  PSR J1910$-$59 is rather weak and would
not be easily detectable using conventional `non-accelerated' search codes.  

All four of the associated clusters are relatively nearby, have high central
luminosity densities L$_{0}$ $\sim 10^5$ L$_\odot$ pc$^{-3}$, and
concentrated cores.  NGC 6397 and NGC 6544 lie within the top four places of
a list of globular clusters (\cite{har96}) ordered by either distance from
the Sun or central luminosity density. 

\subsection{ PSR J1701$-$30 in NGC 6266}
PSR J1701$-$30 has the longest pulse period of the four pulsars, 5.24 ms. 
The orbital period, 3.8 days, is also the longest of the four, and the mass 
function gives a minimum companion mass of 0.19 M$_\odot$. The parameters in 
Table 1 are derived from a fit to 17 independent measurements of the 
apparent pulse period between 1999 December and 2000 July. This system 
is typical of many low-mass binary pulsars, either associated with globular 
clusters or in the Galactic disk (Camilo et al. 2000\nocite{clf+00}). 
Its companion is probably a low-mass 
helium white dwarf. The host cluster, NGC 6266, is listed in the catalogue 
as having a collapsed core, although this is not certain (\cite{djo93}).

\subsection{PSR J1740$-$53 in NGC 6397}

NGC 6397 is a prime candidate for globular cluster pulsar searches (e.g.,
Edmonds et al. 1999\nocite{egc+99}). As mentioned above, it is close and has
a very dense and probably collapsed core (\cite{ksc95}).  Between 2000 July
20 and July 28, PSR J1740$-$53 was observed on 26 occasions but only
detected on about half of these. The observed variations in apparent pulsar
period over the July observing session allowed determination of the orbital
parameters listed in Table 1. When the observed signal strength was plotted
against orbital phase (Fig. 2), it became clear that the pulsar was not seen
at orbital phases between about 0.05 and 0.45.  These non-detections are
centered on orbital phase 0.25 when the pulsar is most distant from the
Earth and hence most likely to be eclipsed by a wind emanating from the
companion. Similar eclipses have been seen in other binary systems, for
example, PSR B1957+20 (Fruchter et al. 1990\nocite{fbb+90}), PSR B1744$-$24A
in the cluster Terzan 5 (Lyne et al. 1990\nocite{lmd+90}; \cite{nt92}) and
PSR J2051$-$0827 (Stappers et al. 1996\nocite{sbl+96}). All of these are
very close binary systems with orbital periods of just a few hours and very
light companions (minimum mass $<$ 0.1 M$_\odot$).  Creation of the
eclipsing wind by ablation of the companion by energetic particles from the
pulsar is a viable mechanism in these cases (\cite{rst91}; Thompson et
al. 1994\nocite{tbep94}).  

\epsfxsize=7.5truecm
\epsfysize=7.0truecm
\epsfbox{fig2.ps}
\vskip -0.5truecm
\figcaption[fig2.ps]{\label{fig:fig2} 
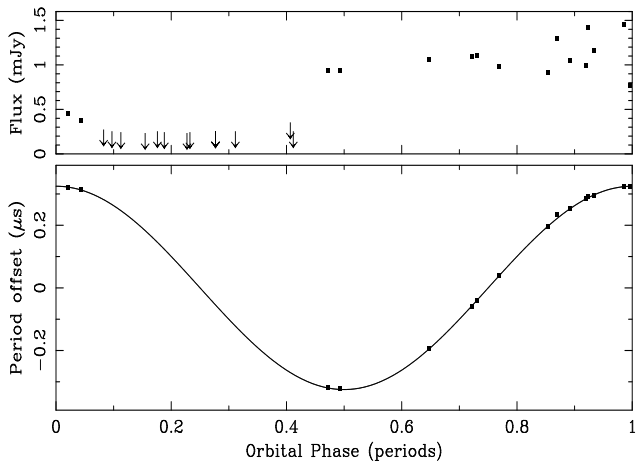
\small{Observed offsets in the apparent pulse period of 
PSR J1740$-$53 in NGC 6397
(corrected to the solar-system barycenter) from the mean value of 3.650330 ms,
and observed flux values through the orbital period. Each point
corresponds to an observation
of 2100 sec duration. The arrows in the upper plot indicate upper limits from 
observations
where the pulsar was not detected.}}
\medskip

In contrast, PSR J1740$-$53 is in a rather wide
binary orbit of period 1.35 days. Assuming a typical surface dipole magnetic
field of $5\times 10^8$ G for the pulsar, the resulting energy loss rate
$\dot{E}_{35}\sim 0.5$ (where $\dot{E}_{35}$ is the spin-down luminosity in
units of $10^{35}$ erg s$^{-1}$) implies a pulsar wind energy density
impinging on
the companion significantly less than that estimated for other
close eclipsing systems. It seems unlikely that a wind of sufficient
density could be driven off a degenerate companion, because the optical
depth at 1.4 GHz for free-free absorption $\tau_{ff}$ scales as the 7$^{\rm
th}$ power of the orbital separation (Rasio, Shapiro, \& Teukolsky 1991).

Another possibility is that the orbital inclination is rather small, $\sim
20^\circ$, and the companion is a normal star of mass comparable to the
turn-off mass of the cluster, $\sim 0.8~{\rm M_\odot}$. Although such a star
(of radius $\gsim 0.8 {\rm R_\odot}$) is well inside its Roche Lobe
($R_{RL}=2.2 {\rm R_\odot}$), it presents a much larger cross-section and
has a more loosely bound atmosphere than a lighter degenerate companion,
favoring the release of a wind sufficiently dense to produce the eclipse
(\cite{pod91}). In this case, for low wind temperatures, $T\sim 10^4$ K,
free-free absorption alone could be a viable eclipse mechanism with
$\tau_{ff} \sim 2\times 10^3~T_4^{-3/2}v_8^{-6} (f~\dot{E}_{35})^2,$ where
$T_4$ and $v_8$ are the temperature and the wind terminal velocity in units
of $10^4$ K and $10^8$ cm s$^{-1}$ and $f$ represents the efficiency of
conversion
of the pulsar spin-down energy in an outflow of matter from the
companion. For hotter winds ($T\gsim 10^5$ K) other absorption processes
may need to be invoked (Thompson et al. 1994).

\subsection{PSR J1807$-$24 in NGC 6544}
NGC 6544 is also one of the closest, densest and most concentrated globular
clusters known.  A radio continuum source of flux density 1.2 mJy was
discovered at its center by Fruchter \& Goss (2000)\nocite{fg00}, but up to
now, pulsar searches have been unsuccessful.  PSR J1807$-$24 is the
strongest of the four pulsars with a mean flux density at 1374 MHz of 1.3
mJy. This is very similar to the flux density of the radio continuum source
and it is almost certain that most or all of the radio flux from this source
is from the pulsar. We have assumed the continuum source position,
R.A.(J2000) =$18^{\rm h} 07^{\rm m} 20\fs24$, Dec.(J2000) =$-24^\circ
59^\prime 25\farcs3$, when deriving the pulsar spin and orbital parameters
given in Table 1. This position is about 26 arcsec from the nominal cluster
center.

Timing observations of the pulsar were conducted using the 76-m Lovell
 Telescope at Jodrell Bank Observatory at a central frequency of 1396 MHz.
 A coherent fit to a total of 99 pulse times of arrival over a 7-day period
 starting on 2000 July 7 gave timing residuals with an rms of 94 $\mu$s.  The
 resulting parameters are given in Table 1 which shows that the orbital
 period is extremely short, 1.7 hours, the second shortest known. Even more
 interestingly, the projected semi-major axis of the orbit is tiny, only 12
 light-ms. The corresponding minimum companion mass is only 0.009 M$_\odot$
 or about 10 Jupiter masses. Apart from the planetary system around PSR
 B1257+12 (\cite{wf92}; \cite{wol94}), this is the least massive pulsar
 companion known. There is no evidence for any eclipse of PSR J1807$-$24 or
 significant dispersive delay, unlike many other pulsars with low-mass
 companions which are eclipsed by a wind from the companion during part of
 the orbit.  This is somewhat surprising as the pulsar and companion are
 only about 500,000 km apart and ablation by the pulsar radiation might be
 expected to be strong.  One possibility is that the system has a small
 inclination angle so that it is more face-on. In this event the companion
 mass would be somewhat larger than the minimum value and hence the system
 may be similar to eclipsing systems such as PSR B1957+20.  Alternatively,
 the companion may in fact be very light and of a different structure or
 composition to those of the eclipsing systems, for example, a low-mass
 brown dwarf or massive planet. This pulsar was independently discovered by
 Ransom et al. (2000)\nocite{rgh+00}.

\subsection{PSR J1910$-$59 in NGC 6752}
NGC 6752 is believed to have a collapsed core although it is less centrally
concentrated than NGC 6397 (Ferraro et al. 1997\nocite{fcb+97}).  The
cluster also has a large proportion of binary systems in its core
(\cite{rb97}).  PSR J1910$-$59 was discovered in 4 consecutive data sets of
length 2100 sec, recorded at Parkes on 1999 October 17, showing a
significant acceleration ($\sim$ --2.2 m s$^{-2}$) on each data set. The
pulsar has a relatively low DM, 34 cm$^{-3}$pc, and it is therefore not
surprising that it scintillates strongly, similar to the pulsars in 47
Tucanae (Camilo et al. 2000\nocite{clf+00}). Because of this and its low
mean flux density (Table 1), it is difficult to detect, with only seven
successful observations so far out of 20. These are insufficient to allow
determination of the binary parameters by fitting to apparent
periods. However, the orbital parameters have been obtained by using the
method recently described by Freire, Kramer, \& Lyne (2000). This primarily
involves fitting an ellipse to observed values of period and acceleration
at different epochs. The orbital period is about 21 hours and the minimum
companion mass is 0.19 M$_\odot$. Apart from the shorter orbital period,
this system is very similar to that in NGC 6266 and is typical of binary
systems with a low-mass helium white dwarf companion.

\section{Implication for the Galactic electron distribution}

Because of their known distances, pulsars in globular clusters provide an
important constraint on the distribution of free electrons in the
interstellar medium. In particular, most are at relatively large
$z$-distances and hence they place an important constraint, in fact
essentially the only constraint, on the `vertical' extent of the Galactic
electron layer. Table 1 lists the parameter DM sin $|b|$, the `vertical'
component of the dispersion measure. The three pulsars at larger
$z$-distances have remarkably similar values of this parameter. These values
are consistent with those of other clusters at similar $z$-distances
(\cite{bv91b}) and indicate a scale height for the electron layer of between
500 and 1000 pc in directions toward the Galactic Center. The remaining
cluster, NGC 6544, lies within the electron layer and has a value for DM sin
$|b|$ consistent with values for other pulsars with similar independently
measured $z$-distances.

\section{Conclusions}
We have discovered short-period binary millisecond pulsars in four globular
clusters which contained no previously known pulsars.  These detections
break the long hiatus in such discoveries and will help in the current
understanding of millisecond pulsars in globular clusters. In fact, the four
clusters presented here are very close and dense, and the absence of
millisecond pulsars in their core was rather intriguing.  The new
discoveries result from the use of enhanced receiving systems having
low system noise, wide bandwidths and good frequency resolution, combined
with the use of high-speed data acquisition systems and improved search
algorithms operating on powerful computing systems. They demonstrate that
many millisecond pulsars, especially those in short-period binary systems,
remain to be discovered in the globular clusters of our Galaxy.  Such
systems provide important constraints on the formation and evolution of
binary systems in clusters and are significant in investigations of cluster
dynamics. At present, it is not at all obvious why some clusters (e.g., 47
Tucanae) have large numbers of detectable pulsars, whereas other apparently
similar clusters (e.g., NGC 6544) have few or none.

\acknowledgements
\small{We thank the staff of Parkes Observatory for their support of this
project
and Ingrid Stairs, Froney Crawford and Gailing Fan for assistance with the
observations. The filterbank system was constructed at Jodrell Bank
Observatory and the Osservatorio Astronomico di Bologna and we especially
thank Tim Ikin for his efforts. FC is supported by NASA grant NAG 5-3229.
The Parkes radio telescope is part of the Australia Telescope which is
funded by the Commonwealth of Australia for operation as a National Facility
managed by CSIRO.}

\end{document}